# A Novel Microwave Dielectric Resonator Based Differential Frequency Sensor for Angular Displacement Detection

A. V. Praveen Kumar[a,*] and Premsai Regalla[a]
[a]EEE Department, BITS-PILANI, Pilani campus, Rajasthan, 333031, India.
* Corresponding author: A.V. Praveen Kumar (Email address: praveen.kumar@pilani.bits-pilani.ac.in)

*Abstract*—A differential frequency microwave sensor for angular displacement detection is reported. A metal strip-loaded cylindrical dielectric resonator (SLCDR) is excited with a 50 Ω microstrip transmission line through a rectangular slot made on its ground plane. Analysis of the transmission coefficient ($|S_{21}|$) of the circuit shows that for the parallel alignment of SLCDR ($\theta = 0^0$), dual-transmission zeros are excited at frequencies $f_L$ and $f_H$, while for the perpendicular alignment ($\theta = 90^0$), a single transmission zero is excited at $f_0$ such that $f_L < f_0 < f_H$. In the range $0^0 \leq \theta \leq 90^0$, frequency $f_L$ increases towards $f_0$ while $f_H$ decreases towards $f_0$. Based on this trend, a differential frequency parameter $\Delta f = f_H - f_L$ is framed as the indicator of the angular displacement following a simulation study with ANSYS HFSS. Subsequent prototype fabrication and VNA measurement confirm the simulations with 15.5 MHz/$^0$ sensitivity over the sensitivity over $90^0$ dynamic range with excellent linearity. Subsequently, using linear inverse regression, the measured differential frequencies are mapped to the angular displacement, and the calculated regression metrics prove very good mapping.

## 1. INTRODUCTION

RF and microwave sensors offer high sensitivity, compact size, and flexibility in terms of circuit design and operation [1]. Microwave sensors exhibit greater robustness and higher precision than low-frequency sensors like ultrasonic, piezoelectric, and resistive types. Additionally, microwave sensors are passive and more cost-effective than high-frequency sensors, such as optical interferometers [1]-[4]. Microwave circuits also offer various characteristic electrical signals or parameters, supporting innovative sensor designs. Recently, microwave sensors have been proposed for displacement detection, a critical process in control and automation industries. The most practical and convenient classification of microwave displacement sensors is based on their operating principles. Specifically, these sensors are categorized into four types such as variable frequency [5]-[15], fixed frequency-variable magnitude [16]-[27], variable phase [28]-[29], and differential sensors [30]-[33]. All of these utilize microwave resonators as sensing elements, which exhibit excellent response to alignment or displacement, leading to changes in the resonant characteristics of the circuit such as resonant frequency, magnitude, phase, and bandwidth. Displacement can be linear, angular, or a combination of both. Microwave resonator-based angular displacement sensors are particularly appealing due to their compact design, flexible operation, high accuracy, and wide detection range [5]-[19]. These attributes make them essential for various scientific and industrial applications, including the detection and correction of rotational movement in control and automation industries and spacecraft systems [1], [18].

Single-ended microwave sensors utilize a single output parameter, such as the resonant frequency, bandwidth, magnitude or phase of the resonant circuit. Each operation provides distinctive advantages and disadvantages with respect to the sensor's performance metrics, such as sensitivity, accuracy, dynamic range, and practical implementation [5]-[33]. Generally, the practical design of a microwave sensor is more straightforward and cost-effective with a single-frequency or narrow-band measurement system compared to a swept-frequency system [18], [19], [23], [24]. The variable-magnitude and variable-phase sensors are examples of single-frequency sensors, providing unique immunity to variations in ambient conditions [23],[24],[28]. When simple implementation is the primary concern, variable-magnitude operation is preferred, whereas extremely high sensitivity requirements favor variable-phase sensing [28], [29]. Variable-frequency sensors are attractive for a wider dynamic range and are immune to amplitude errors caused by cross-coupling and EMI [5], [13]-[15]. However, these suffer from frequency detuning errors resulting from variations in ambient conditions [1]. To fix this accuracy issue, differential frequency sensors have been proposed [30]-[33]. Most of the differential frequency sensors are dedicated to material characterization, particularly for permittivity measurement. In [30], a microstrip line is loaded with a pair of stepped-impedance resonators, each of which is independently perturbed dielectrically. When the two resonators have identical permittivities, the structure displays a single resonance frequency. However, if the permittivities differ, the resonance frequency splits, resulting in two distinct resonances. The separation between these resonance frequencies is utilized to detect permittivity. Similarly, in [31], a symmetry disruption-based CSRR loaded transmission line structure was proposed to detect the permittivity of the sample. In [32], a two-dimensional alignment and displacement sensor is established based on a movable broadside-coupled split ring resonator (BC-SRR). The sensor operates on the principle of structural symmetry, where breaking this symmetry results in splitting the resonance into two distinct notches. The frequency difference between these notches is used to detect displacement. Later, in [33], an H-shaped displacement sensor is established to extend the dynamic range. It consists of coupled lines acting as the stator and an open stub as the mover. As an open stub resonator moves, frequency splitting occurs in the form of even and odd mode resonances, and the difference between both split frequencies is used to detect the displacement. In the literature, sensors based on the differential frequency principle are predominantly limited to linear displacement detection [32],[33]. Therefore, further investigations in this direction are essential for angular displacement detection to exploit the advantages of differential frequency sensing, such as wider dynamic range, higher sensitivity, and higher accuracy, compared to single-frequency or swept-frequency sensors [5]-[27].

To realize a differential frequency sensor using the microwave technique, the circuit must first generate dual resonances either by employing two separate resonators or operating a single resonator in dual resonant modes. Like metallic resonators, dielectric resonators (DRs) are versatile candidates for various applications due to their inherent features like mode diversity, excitation feasibility, and degree of design freedom [34], [35]. Previously, DRs were implemented in various sensing applications in the environmental, chemical, civil, and biomedical fields [36]-[40]. The scope has recently been expanded to displacement sensors as well [18], [19], [22]–[24]. All the above-mentioned DR-based linear and angular displacement sensors have been developed based on the magnitude variation ($|S_{11}|$ or $|S_{21}|$) principle at a fixed frequency. Notably, a single DR unit can support multiple resonant modes that can be excited with common microwave feeding methods [42], [43]. Hence in the proposed work, a DR-based differential frequency microwave sensor is developed for the first time to detect the angular displacement in the $0^0$ to $90^0$ range. Unlike the previously reported differential sensors [32], the proposed sensor uses a single resonator unit to excite dual resonances. The proposed sensor exhibits high sensitivity, excellent linearity, and a wide dynamic range, as validated through HFSS simulation, VNA measurement, and inverse regression analysis. The remainder of the paper is organized as follows. Section 2 covers sensor conceptualization and design. Section 3 provides a simulation analysis. Section 4 deals with the prototype fabrication and experimental results, followed by regression analysis in Section 5. Lastly, Section 6 discusses the advantages of the proposed sensor in comparison to existing angular sensors, with relevant conclusions summarized in Section 7.

## 2. CONCEPTUALIZATION AND DESIGN OF THE SENSOR

### 2.1 Conceptualization

Developing the differential frequency sensor fundamentally requires the excitation of at least two resonances. This can be achieved either by simultaneously exciting the resonant modes of two separate resonators with the same feed [43] or by using a single resonator excited in dual modes [42]. For instance, Fig. 1 shows the ideal transmission response of a dual-mode resonator, giving rise to two transmission zeros (TZs) at $f_L$ and $f_H$. Suppose that any change in the geometrical symmetry of the resonator due to the angular alignment of the resonator modifies the resonant frequencies in such a way that $f_L$ increases and $f_H$ decreases with the angle. The resulting differential frequency $\Delta f = f_H - f_L$ provides a measurable output quantity which can be mapped to the corresponding angular displacement ($\theta$), indicating a sensitivity $S = \Delta f / \Delta \theta$. The higher the frequency separation, the higher the sensitivity and the more uniform the frequency shift with angle, the more constant the sensitivity.

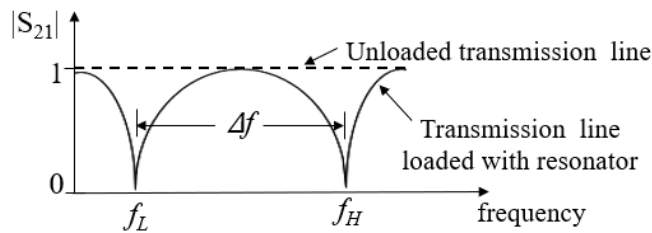

Fig.1. Conceptual transmission spectrum for the proposed differential frequency sensor.

### 2.2 Sensor design model

In the present work, the conceptual transmission spectrum (Fig.1) is practically achieved by using a two-port 50 Ω microstrip transmission line exciting a cylindrical dielectric resonator (CDR) through a narrow slot on the ground plane. The slot coupling is an effective scheme to excite multiple resonating modes of a CDR [42]. The resulting structure excites multiple resonant modes, of which the lower two modes have been considered for generating the differential frequency. The slot coupling structure made on the dielectric substrate forms the stationary part of the angular sensor, while the CDR forms its rotary part to which the angular displacement is applied. The sensor design involves four stages as illustrated in Fig.2(a)–(d). Corresponding design parameters are specified in Table 1. Note that the substrate and the CDR parameters are kept constant due to concerns about availability.

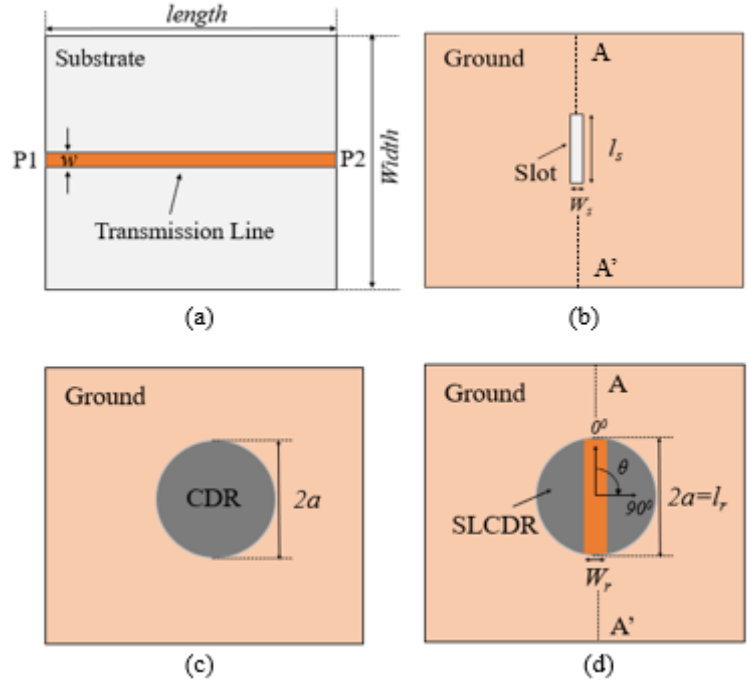

Fig.2. HFSS simulation model of the proposed sensor (a) substrate side with the transmission line, (b) ground plane side with the slot, (c) CDR loaded on the slot, and (d) SLCDR loaded on the slot

**Table 1. Design parameter specifications of the proposed sensor**

| Involved parameters | Values |
|---|---|
| *Dielectric substrate* | |
| Relative permittivity ($\varepsilon_{r1}$) | 4 |
| Loss tangent (tan$\delta$) | 0.02 |
| Height | 1.6 mm |
| Length × width | $100 \times 100$ mm$^2$ |
| *50 Ω Transmission line and slot* | |
| Transmission line width | 3.2 mm |
| Slot length ($l_s$) × width ($w_s$) | $12 \times 2$ mm$^2$ |
| *Cylindrical Dielectric Resonator* | |
| Diameter ($2a$) | 19.43 mm |
| Aspect ratio (a/d) | 1.33 |
| Relative permittivity ($\varepsilon_{r2}$) | 24 |
| Loading metal strip's length ($l_r$) × width ($w_r$) | $19.43 \times 3.2$ mm$^2$ |

High-frequency modelling and simulations of the structure are performed by using ANSYS HFSS [41], and the transmission spectrum ($|S_{21}|$) is generated. It shows two distinct TZs at $f_L$ and $f_H$, which, from the near-field patterns of the CDR, are confirmed as the HEM$_{11\delta}$ mode of the CDR ($f_L$) and the CDR loaded slot mode ($f_H$) [42]. By keeping both the substrate and CDR parameters fixed, the slot size is varied to study how the resonances are affected, and the results are presented in Table 2. The results reveal that $l_s$=12 mm provides optimal design, favoring the frequency separation ($\Delta f$) and corresponding notch levels ($|S_{21}|_{min}$). The corresponding $|S_{21}|$ spectrum is shown in Fig. 3, where $f_L$ =2.52 GHz and $f_H$ =4.1 GHz are marked.

**Table 2. Resonant frequencies and corresponding |S$_{21}$| values of a simple CDR with the various slot lengths ($l_s$)**

| $l_s$ (mm) | frequency (GHz) | | $\Delta f = f_H - f_L$ | $|S_{21}|_{min}$ (dB) | |
|---|---|---|---|---|---|
| | 1$^{st}$ TZ ($f_L$) | 2$^{nd}$ TZ ($f_H$) | | 1$^{st}$ TZ | 2$^{nd}$ TZ |
| 6 | 3.02 | 5.18 | 2.16 | −2.38 | −10.51 |
| 8 | 2.91 | 4.7 | 1.79 | −5.62 | −14.91 |
| 10 | 2.75 | 4.29 | 1.54 | −10.86 | −14.56 |
| 12 | 2.52 | 4.1 | 1.58 | −16.28 | −13.4 |
| 14 | 2.35 | 3.94 | 1.59 | −19.93 | −11.83 |
| 16 | 2.16 | 3.87 | 1.71 | −22.81 | −10.71 |

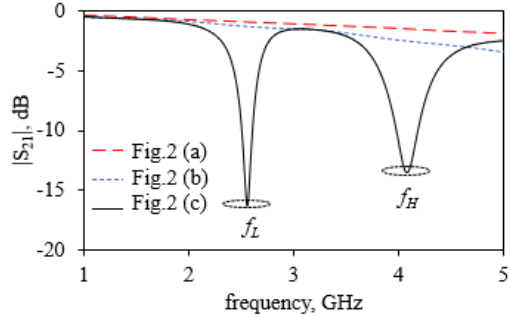

Fig.3. Simulated $|S_{21}|$ spectrum of the proposed sensor for each design stage, as shown in Fig.2. (a) − (c).

### 2.3 Enabling angular sensitivity

As the axisymmetric CDR can't introduce angular sensitivity, a rectangular metal strip is affixed to the top face of the CDR. The dimensions of the metal strip are shown in Table 1. This modified resonator is called the strip-loaded cylindrical dielectric resonator (SLCDR). The slot direction marked A-A' in Fig. 2(b) and (d), is the reference axis for indicating angular displacement where $\theta = 0^0$ implies parallel alignment of the strip relative to the slot while for $\theta = 90^0$, the alignment is orthogonal. Corresponding $|S_{21}|$ responses for various cases are plotted in Fig.4, which shows that for the SLCDR with $\theta = 0^0$, the TZs appear at 2.5 GHz and 4.02 GHz, almost identical to that of the simple CDR. The same figure also shows that, for $\theta = 90^0$, i.e., the orthogonal alignment between the strip and the slot, there is only one TZ at 3.1 GHz. Due to the structural symmetry, the angular sensitivity is limited to $90^0$.

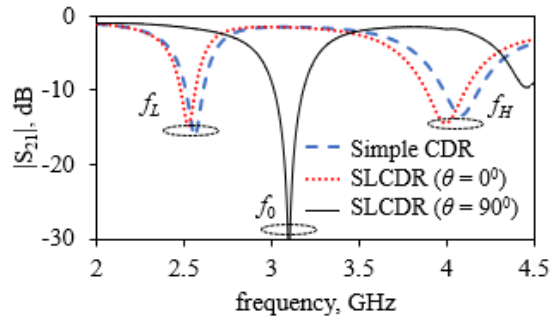

Fig.4. Simulated $|S_{21}|$ spectrum of the proposed sensor for $\theta = 0^0$ and $90^0$ for the CDR and SLCDR

### 3. SIMULATED ANGULAR SENSITIVITY

The angular variation of the TZs of the proposed sensor across $0^0$ to $90^0$ is illustrated in the $|S_{21}|$ spectrum of Fig. 5 and the corresponding variation in the individual frequencies in Fig.6. Figures clearly show that as $\theta$ increases, $f_L$ increases while $f_H$ decreases, finally converging to $f_0$ at $90^0$.

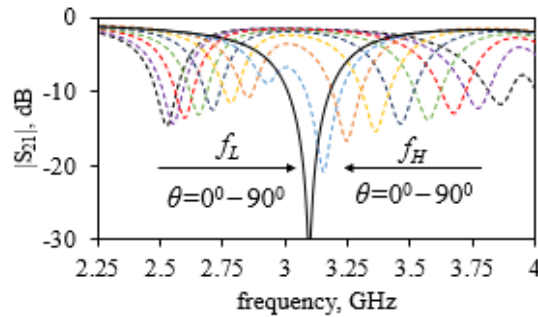

Fig.5. Simulated $|S_{21}|$ vs. frequency response of the proposed sensor for change in $\theta$

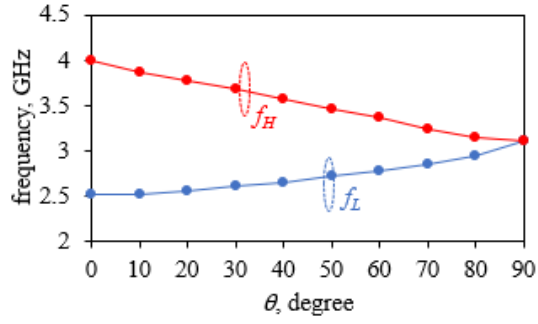

Fig.6. Simulated variation in $f_L$ and $f_H$ with $\theta$ of the proposed sensor

In Fig.6, the frequency variation of $f_L$ can be noted from 2.5 GHz to 3.1 GHz, whereas that of $f_H$ is from 4 GHz to 3.1 GHz. Fig. 7 illustrates the simulated $\Delta f$ vs. $\theta$ response of the proposed sensor giving a frequency span of 1500 MHz over $90^0$, implying an average sensitivity of 16.7 MHz/$^0$ according to eq. (1).

$$Sensitivity = \frac{\Delta f}{\Delta \theta} \quad (1)$$

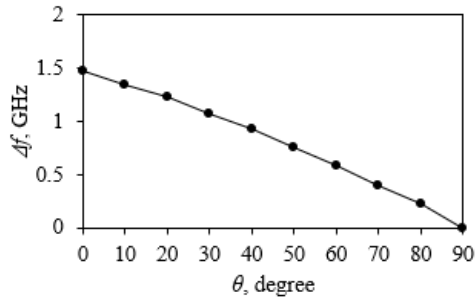

Fig.7. Simulated $\Delta f$ vs. $\theta$ response of the proposed sensor

The opposing senses of angular dependences of $f_L$ and $f_H$ shown in Fig. 6 can be explained by referring to the near-field distributions of the SLCDR shown in Fig. 8, generated in HFSS. According to Fig.8(a) and (c), at 2.5 GHz, the E-field is stronger near the upper CDR face, whereas as per Fig.8(b) and (d), at 4.02 GHz, it is the H-field that is dominant near the upper face. For $\theta = 0^0$, the strip alignment is such that it short circuits the E-field of the CDR mode, causing maximum electric field perturbation, while for $\theta = 90^0$, the strip couples maximally to the slot mode, perturbing its magnetic field. According to the perturbation theorem [44], $f_L$ increases from 2.5 GHz while $f_H$ decreases from 4.02 GHz between the two extremities, as the $\theta$ increases. At $\theta = 90^0$, both the TZs merge at $f_0 = 3.1$ GHz with the near-field patterns as shown in Fig.9.

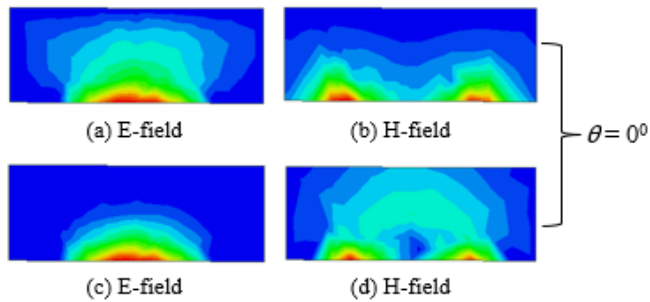

Fig.8. Field distributions for the SLCDR ($\theta = 0^0$) : (a) & (b) at $f_L = 2.5$ GHz, (c) & (d) at $f_H = 4.02$ GHz

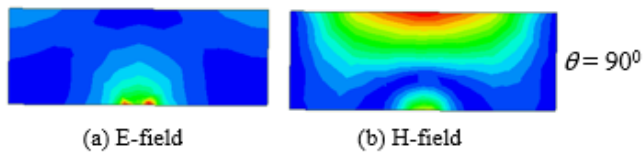

Fig.9. Field distributions for the SLCDR ($\theta = 90^0$) at 3.1 GHz

## 4. SENSOR FABRICATION AND TESTING

A prototype model of the proposed sensor is fabricated and tested experimentally to verify the sensor concept and simulated results. The specifications of the fabricated prototype closely resemble those of the simulated model, as in Table 1. Fig. 10 displays a photograph of the fabricated prototype corresponding to Fig. 2. An acrylic-based, non-metallic support structure is designed to facilitate precise angular movements and reduce proximity effects such as frequency detuning. Measurements are conducted by using Keysight Technologies N9928A Vector Network analyzer (VNA), as illustrated in Fig. 11.

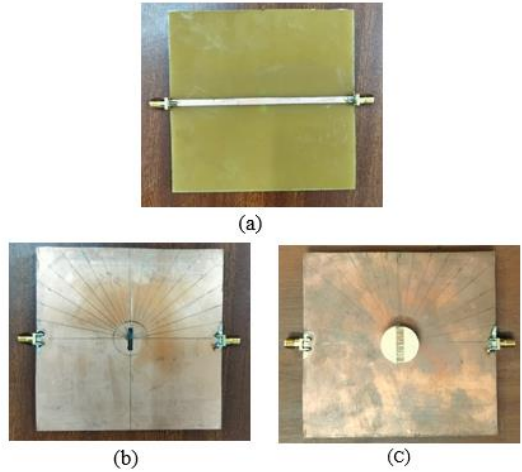

Fig.10. Photograph of the fabricated prototype of the proposed sensor (a) substrate side with the transmission line, (b) ground plane side with the slot (c) SLCDR loaded on the slot (All the dimensions are identical to the HFSS model)

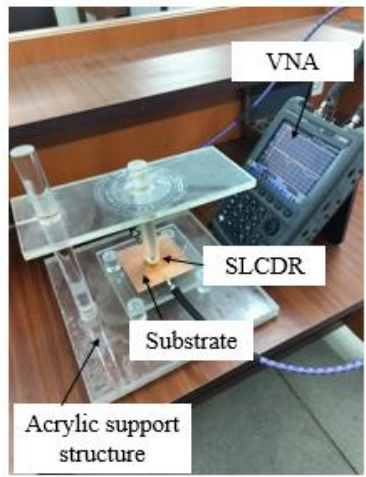

Fig.11. Sensor prototype measurement setup

Full 2-port calibration of the VNA with port extensions (coaxial cables) is performed using the standard open, short, and matched loads available with the 85052D Cal kit over the desired frequency range with 801 frequency points to achieve high accuracy. The calibration effectively eliminates various systematic and drift errors. Fig. 12 shows the $|S_{21}|$ vs $\theta$ response measured by the VNA. At $\theta = 0^0$, the measured transmission zeros were observed at 2.52 GHz and 3.92 GHz, while the simulated values occurred at 2.5 GHz and 4.02 GHz. At $\theta = 90^0$, the measured transmission zero occurred at 3.18 GHz, while the simulated value was at 3.1 GHz. Fig. 13 shows the measured frequency variation response for both the $f_L$ and $f_H$ and is compared with the simulated response. Fig. 14 presents the measured $\Delta f$ response compared with the simulation results. The measured average sensitivity is 15.5 MHz/$^0$ for $90^0$ range. Measurements were performed three times to ensure repeatability, and the average values are reported. Fig. 14 clearly shows that the sensor's response is highly linear, implying nearly constant sensitivity. Therefore, in the next section, a linear inverse regression model is applied to estimate the angular displacement from $\Delta f$ measurement.

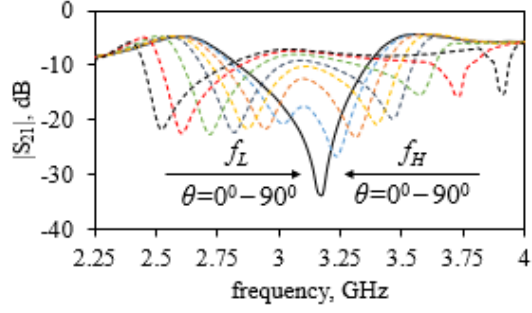

Fig.12. VNA measured |S$_{21}$| vs. frequency response of the proposed sensor for change in $\theta$

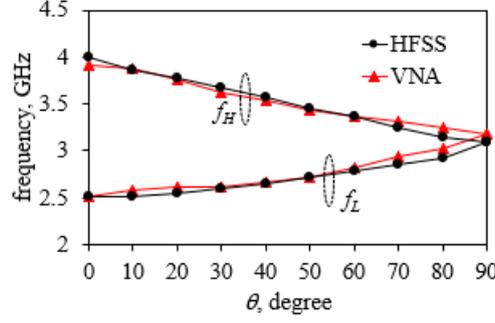

Fig.13. Simulated (HFSS) and measured (VNA) variation in $f_L$ and $f_H$ with $\theta$ of the proposed sensor

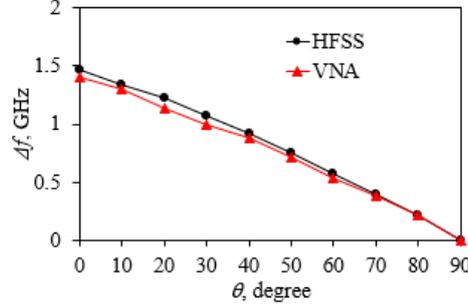

Fig.14. Simulated (HFSS) and measured (VNA) $\Delta f$ vs. $\theta$ response of the proposed sensor

## 5. INVERSE REGRESSION ANALYSIS

To estimate the angular displacement ($\theta$) from measured $\Delta f$, a linear inverse regression model is used, which is analytically expressed as [45],

$$\theta = A \times \Delta f + B \quad (2)$$

Where $\theta$ is the response variable or regressand, $\Delta f$ is the observed variable or regressor, $A$ is the gradient, and $B$ is the intercept. For the proposed model, the achieved gradient and intercepts are −64.93 and 94.22. Fig. 15. (a) shows the estimated displacement values vs. the actual measured displacement values. The regression metrics like estimation error, root mean square error (RMSE), and R-squared value are calculated to validate the model, using the following eqs. (3)-(6),

$$Error(^0) = Predicted\_\theta - Actual\_\theta \quad (3)$$

$$Error\ (\%) = \frac{Predicted\_\theta - Actual\_\theta}{Actual\_\theta} \times 100 \quad (4)$$

$$RMSE = \sqrt{\frac{\sum_i^n (Predicted\_\theta_i - Actual\_\theta_i)^2}{n}} \quad (5)$$

$$R^2 = 1 - \frac{\sum_i^n (Predicted\_\theta_i - Actual\_\theta_i)^2}{\sum_i^n (Predicted\_\theta_i - \overline{Actual\_\theta})^2} \quad (6)$$

Where, $Predicted\_\theta_i$ is the predicted $\theta$ value for the $i^{th}$ data point, $Actual\_\theta_i$ is the measured $\theta$ value at $i^{th}$ data point, and $\overline{Actual\_\theta}$ is the mean of the actual data. By following the eq. (3), as shown in Fig. 15. (b), the acheived maximum error is $3.8^0$ at $90^0$, whereas by following eq. (4), the maximum error percentage obtained was 7.3 % at $40^0$. According to eq. (5), the calculated root mean square error (RMSE) is $2.13^0$ or 3.61 %. The corresponding R-squared value was calculated using eq. (6) is 0.997, showing an excellent fit to the regression model [46]. The achieved sensor metrics are highlighted in Table. 3.

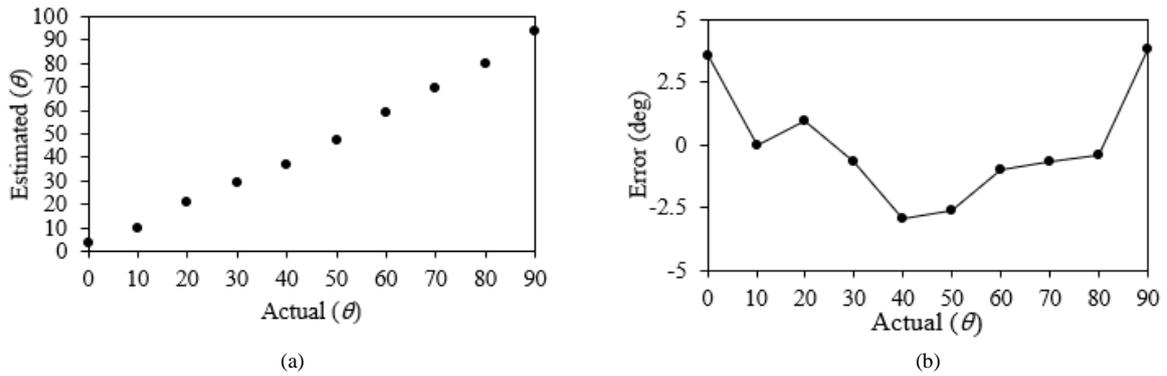

(a) (b)
Fig.15. Angular displacement estimation using inverse regression model (a) estimated vs actual displacement, (b) estimation error

Table 3. Regression and error analysis results of the proposed sensor

| Regression type | Linear |
|---|---|
| Input Range | $0^0$ to $90^0$ |
| $R^2$ Value | 0.997 |
| Max error ($^0$) | 3.8 at $90^0$ |
| Max error (%) | 7.3 at $40^0$ |
| RMS Error ($^0$) | 2.13 |
| RMS Error (%) | 3.61 |

Table 4.
Comparison of the proposed sensor's performance with existing variable frequency microwave sensors for angular displacement detection

| [Ref] Year | Sensing element | Freq. range $f_L$/$f_H$ GHz | Sensitivity (MHz/$^0$) | Dynamic range ($^0$) | FOM (Sensitivity × Dynamic range) | Linearity | Displacement estimation using inverse regression analysis |
|---|---|---|---|---|---|---|---|
| [5] 2014 | U-shaped resonator | 0.87/1.2 | 1.85 | 180 | 333 | Not linear | No |
| [6] 2018 | NSRR | 1/1.58 | 14.4 | 40 | 576 | Moderately linear | No |
| [7] 2019 | CSRR | 5.61/5.83 | 2.37 | 90 | 213.3 | Not linear | No |
| [9] 2019 | ML TSI | 0.86/1.1 | 1.24 | 180 | 223.2 | Not linear | No |
| [10] 2020 | MLTSI | 0.65/1.2 | 3.15 | 180 | 567 | Moderately linear | No |
| [11] 2020 | ML | 3.56/4.38 | 4.56 | 180 | 820.8 | Not linear | No |
| [12] 2021 | TFS | 2.35/2.15 | 2.27 | 90 | 204.3 | Not linear | No |
| [13] 2021 | ML RCSR | 0.82/1.26 | 1.22 | 360 | 439.2 | Not linear | No |
| [14] 2022 | CSRR | 2.39/6.34 | 21.94 | 180 | 3949.2 | Not linear | No |
| [15] 2023 | ML CSR | 2.12/0.88 | 4.03 | 310 | 1249 | Not linear | No |
| This work | SLCDR | 2.52 / 4 | 15.5 | 90 | 1395 | Highly linear | Yes |

## 6. COMPARISON WITH EXISTING WORKS

To manifest the usefulness of the proposed differential frequency angular sensor, a comparison with the existing swept frequency sensors is made and summarized in Table. 4, as reported works on differential frequency angular sensors are not available in the existing literature. In Table. 4, various performance indicators such as frequency range, sensitivity, linearity, dynamic range, figure-of-merit (FOM), and displacement estimation metrics are compared. The proposed sensor exhibits excellent linearity, enhanced sensitivity, and a superior figure-of-merit (FOM) over $90^0$ range. Compared to [14], the present sensor does not require any additional lumped elements to improve the quality factor or vias through the ground to form a resonator. Unlike the sensor in [5], the proposed design is less sensitive to manufacturing tolerances, less prone to undesired radiation effects, and enables the creation of deeper notches (over 20 dB TZ in this work compared to 10 dB in [5]), potentially leading to a more accurate sensing process.

## 7. CONCLUSION

This work presented the design and validation of an SLCDR-based differential frequency sensor to detect angular displacement from $0^0$ to $90^0$. The SLCDR excited by a 50 Ω microstrip line through a rectangular slot gives rise to two distinct transmission zeros in the frequency spectrum from the CDR and the slot contributions. The difference between these resonance frequencies varies with the angular alignment of the SLCDR relative to the slot's fixed position. The sensor design and analysis have been conducted with the help of HFSS simulations, followed by prototype fabrication and laboratory measurement. Regression analysis of the measured data displays decent mapping between the sensor's input and output quantities. The sensor exhibits a measured sensitivity of 15.5 MHz/$^0$ in the $90^0$ dynamic range. The sensitivity can be further enhanced by increasing the frequency difference between the two transmission zeroes through the proper optimizations of the CDR and slot. This will be explored in future work.

ACKNOWLEDGEMENT

This work was supported by the Department of Science and Technology (DST), Govt. of India, through the DST-FIST grant Ref: SR/FST/ETI-376/2013. Also, the authors are thankful to Mr. Mahesh Chandra Saini (Technical assistant, EEE Department, BITS Pilani, Pilani campus, Rajasthan, India) for his assistance in prototype fabrication.